\documentclass[11pt,a4paper]{article}
\usepackage{epsfig,graphicx,amsfonts,amsmath}
\usepackage{cite}
\RequirePackage{mathrsfs}

\newlength{\dinwidth}
\newlength{\dinmargin}
\def\eq#1{{Eq.~(\ref{#1})}}
\newcommand{\Le}{\left(}
\newcommand{\Ra}{\right)}

\newcommand{\tldv}{\tilde{\textsl{v}}}
\newcommand{\beq}{\begin{equation}}
\newcommand{\eeq}{\end{equation}}
\newcommand{\beqar}{\begin{eqnarray}}
\newcommand{\eeqar}{\end{eqnarray}}
\newcommand{\D}{\partial}

\date{}
\begin{document}

\title {{~}\\
{\Large \bf  Effective action for reggeized gluons, classical gluon field of relativistic color charge and color glass condensate approach  }\\}
\author{ 
{~}\\
{\large 
S.~Bondarenko$^{(1) }$,
L.~Lipatov$^{(2,3)}$,
A.~Prygarin$^{(1) }$ 
}\\[7mm]
{\it\normalsize  $^{(1) }$ Physics Department, Ariel University, Ariel, Israel}\\
{\it\normalsize  $^{(2) }$ St.Petersburg Nuclear Physics Institute, Russia}\\
{\it\normalsize  $^{(3) }$ II Institute of Theoretical Physics, Hamburg University, Germany}\\
}
%

\maketitle
\thispagestyle{empty}

\begin{abstract}
 We discuss  application of formalism of small-$x$ effective action for  reggeized gluons, \cite{Gribov,LipatovEff,BFKL}, for the calculation of classical gluon field of relativistic color charge, similarly to that done in CGC approach of \cite{Venug,Kovner}.
The equations of motion with the reggeon fields are solved in  LO and NLO approximations and new solutions are found. 
The results are compared to the calculations performed in the CGC 
framework and it is demonstrated that the LO CGC results for the classical field are reproduced in our calculations. 
Possible applications of the NLO solution  in the effective action and CGC frameworks are discussed as well.

\end{abstract}

\section{Introduction}

 In the framework of perturbative QCD, the calculations of classical field created by a relativistic color charge is an important task and its results are useful in 
many physical applications. There are self-consistent approaches for the calculation of the mean field in the framework of Wong's equations, in the theory
of non-Abelian plasma physics and in the classical and quantum transport theories, see \cite{Heinz,VenPlas,VenPlas1,Iancu,Plas1} and references therein. 
In the context of the Color Glass Condensate (CGC) approach, the knowledge of solution of the equations of motion is an important initial set up for the  further small-$x$ evolution
of the gluonic system, see \cite{Venug,Kovner}.

 The CGC approach is based on the renormalization group (quantum evolution) applied to some initial classical configuration of the
gluon field. Equations of motion for the gluon field in this case are derived  within some effective action, see details in \cite{Venug,Kovner}. 
An important ingredient of the framework 
is a source of the gluon field, see \cite{Kovner} and discussion further in the paper. 
The structure of this term, in turn, determines the form of equations of motion and correspondingly
the form of solution of the equations, which are known to the LO precision. The Balitsky, Fadin, Kuraev, Lipatov
(BFKL) like, \cite{BFKL}, small-$x$ behavior of the gluon density,
a non-linear Gribov, Levin, Ryskin (GLR) equation, see \cite{GLR} as well as Balitsky-Kovchegov (BK) like, \cite{BK}, corrections to this density were reproduced
in the framework of this approach. 

 In the frameworks based on the theory of BFKL Pomeron, \cite{LipatovEff,BFKL}, the source terms in the action are considered as well, 
see \cite{Braun,Braun1,Higgs,Bom,Bond}. There the Pomeron is coupled to the source, it is usual formulation of different 
RFT theories, see \cite{OldRFT,NewRFT1}. The effective action approach of \cite{LipatovEff} can be considered as 
some generalization of Gribov's Regge calculus, \cite{Gribov}, for the case of QCD degrees of freedom. Besides the usual gluon field, it includes
two additional reggeon fields and widely used for the calculations of different quasi-elastic LO and NLO production 
amplitudes in the multi-Regge kinematics, see \cite{EffAct}, or calculation of NLO corrections to the BKP, \cite{BKP}, kernel, see \cite{BKP1}.

 In our paper, we use the effective action for  reggeized gluons exploring ideas of \cite{LipatovEff}. Using light-cone gauge, we consider a problem with
only one longitudinal gluon field in the equations of motion included. With the two reggeon fields present in the approach, 
the first reggeon field is defined as a  LO value of the corresponding gluon field, whereas the second reggeon field
arises as a source term in the Lagrangian. 
The form of the effective  
currents, arising in the equations of motion, therefore, can be obtained or directly from the effective action expression from \cite{LipatovEff} or
from the self-consistency conditions for the solution of the equations of motion, in both cases we obtain the same structure of the current.
Respectively, in the next Section 2, we discuss the set up of the problem, i.e. we consider the expression for the effective action , further, in the Section 3, 
we write equations of motion for the gluon fields of the problem in light-cone gauge. In the Section 4 we solve these equations
with the LO accuracy and  demonstrate that obtained solution   is the same as obtained in CGC approach in the limit of zero reggeon field. Section 5 is dedicated 
to the semi-classical Reggeon action obtained from the initial one and in the Section 6 we discuss
the similarities and differences between the solutions obtained in the paper and in the framework of CGC approach. 
The Conclusion is presented in Section 7, a calculation of the NLO solution is in the Appendix C and in
the  Appendixes A and B
we discuss the form of the current in the effective action for reggeized gluon approach.

\section{Effective action for  reggeized gluons with color field source }

 The effective action, see \cite{LipatovEff}, is a non-linear gauge invariant action which correctly reproduces the production of the particles in direct channels at a quasi-multi-Regge kinematics. 
It is written for the local in rapidity interactions of physical gluons in
direct channels inside of some rapidity interval $(y\,-\,\eta/2,y\,+\,\eta/2)$. In turn, the interaction between the different clusters of gluons at different but close rapidities
can be described with the help of reggeized gluon fields
\footnote{We use the Kogut-Soper convention for the light-cone for the light-cone definitions with $x_{\pm}\,=\,\Le x_{0}\,\pm\,x_{3} \Ra/\sqrt{2}$ and $x_{\pm}\,=\,x^{\mp}$\,.} 
$A_{-}$ and $A_{+}$ interacting in crossing channels. Those interaction are non-local in rapidity space. 
This non-local term is not included in the action, the term of interaction between the reggeon fields in the action is local in rapidity and can be considered as 
some kind of renormalization term  in the Lagrangian.
The action is gauge invariant and written in the covariant form in terms of gluon field $\textsl{v}$ as
\beq\label{EA92}
S_{eff}\,=\,-\,\int\,d^{4}\,x\,\Le\,\frac{1}{4}\,G_{\mu \nu}^{a}\,G^{\mu \nu}_{a}\, \,+\,tr\,\left[\,\Le\,A_{+}(\textsl{v}_{+})\,-\,A_{+}\,\Ra\,j_{reg}^{+}\,+\,
\Le\,A_{-}(\textsl{v}_{-})\,-\,A_{-}\,\Ra\,j_{reg}^{-}\,\right]\,\Ra\,,
\eeq
where
\beq\label{EA921}
A_{\pm}(\textsl{v}_{\pm})\,=\,\frac{1}{g}\,\D_{\pm}\,O(x^{\pm}, \textsl{v}_{\pm})\,=\,
\textsl{v}_{\pm}\,O(x^{\pm}, \textsl{v}_{\pm})\,,
\eeq
with $O(x^{\pm}, \textsl{v}_{\pm})$ as some operators, see \cite{LipatovEff},
Appendix A, Appendix B and Section 4 in  Iancu et al. in \cite{Kovner}.
The form of reggeon current we take is the following\footnote{It is rewritten in comparison to the current 
from \cite{LipatovEff}}: 
\beq\label{EA9233}
j_{reg\,a}^{\pm}\,=\,\frac{1}{C(R)}\,\D_{i}^{2}\,A_{a}^{\pm}\,,
\eeq
where $C(R)$ is the eigenvalue of Casimir operator in the representation R, $C(R)\,=\,N$ in the case of adjoint representation used in the paper.
There are additional kinematical constraints for the reggeon fields
\beq\label{EA911}
\partial_{-}\,A_{+}\,=\,\partial_{+}\,A_{-}\,=\,0\,,
\eeq
corresponding to the strong-ordering Sudakov components in the multi-Regge kinematics,
see \cite{LipatovEff}. Here, as usual,  $\,\partial_{i}\,$ denotes the derivative on transverse coordinates.
In the framework with an external  source of the color charge introduced, 
keeping only gluon field depending terms in the \eq{EA92} action, we rewrite \eq{EA92} as
\beq\label{EA91}
S_{eff}\,=\,-\,\int\,d^{4}\,x\,\Le\,\frac{1}{4}\,G_{\mu \nu}^{a}\,G^{\mu \nu}_{a}\,\,+\,
\textsl{v}_{-}\,J^{-}(\textsl{v}_{-})\,\,+\,
\textsl{v}_{+}\,J^{+}(\textsl{v}_{+})\,\Ra\,,
\eeq
with
\beq\label{Vst1}
J^{\pm}(\textsl{v}_{\pm})\,=\,O(\textsl{v}_{\pm})\,j_{reg}^{\pm}\,.
\eeq
Under variation on the gluon fields these currents reproduce the Lipatov's induced currents
\beq\label{EA9111}
\delta\,\Le\,\textsl{v}_{\pm}\, J^{\pm}(\textsl{v}_{\pm})\,\Ra\,=\,\Le \delta\,\textsl{v}_{\pm} \Ra\,j^{ind}_{\mp}(\textsl{v}_{\pm})\,=\,
\Le  \delta\,\textsl{v}_{\pm} \Ra \,j^{\pm}(\textsl{v}_{\pm})\,,
\eeq
with shortness notation $j^{ind}_{\mp}\,=\,j^{\pm}$ introduced.
This current posseses a covariant conservation property:
\beq\label{EA924}
\Le\,D_{\pm}\,j_{\mp}^{ind}(\textsl{v}_{\pm})\,\Ra^{a}\,=\,\Le D_{\pm}\,j^{\pm}(\textsl{v}_{\pm})\Ra^{a}\,=\,0\,.
\eeq
Here and further we denote the induced current in the component form in the adjoint representation\footnote{We use
$\Le\,T_{a}\,\Ra_{b\,c}\,=\,-\,\imath\,f_{a\,b\,c}$ definition of the matrices and  write only "external" indexes of the $f_{a\,b\,c}\,=\,\Le f_{a}\Ra_{b\,c}$ matrix in the trace. } as 
\beq\label{EA9241}
j_{a}^{\pm}(\textsl{v}_{\pm})\,=\,-\,\imath\,tr[T_{a}\,j^{\pm}(\textsl{v}_{\pm})]\,=\,\frac{1}{N}\,tr\left[\,f_{a}\,O\,f_{b}\,O^{T}\,\right]\,\Le \D_{i}^{2}\,A_{\mp}^{b} \Ra\,,
\eeq
see Appendixes A and B.
It will be shown in the following sections, that the requests of self-consistency of the solution of equations of motion  will lead independently 
to the \eq{Vst1}-\eq{EA924} form and described properties of the current. 
Applying  the light-cone gauge $\textsl{v}_{-}\,=\,0$,  the second current term in the r.h.s. of \eq{EA91} looks equal to zero on the first sight,  but due the
$x_{-}$ derivative in the \eq{EA921} this term acquires some non-zero value on the boundaries of integration, see also discussion in \cite{Hatta1}. For our calculations these contributions are not important, but 
it is interesting to note, that after the 
integration on $x_{\pm}$ variables in the effective currents terms in the Lagrangian, the remaining expressions can be interpreted as distribution of the light-cone straight lines
in the three dimensional space with the probabilities of distribution determined by the reggeon fields. This geometrical picture potentially can be very interesting,
because a deformation of this lines can lead to some interconnections of the approach with framework of \cite{MaMi}.

\section{Equations of motion in light-cone gauge}

 The 
classical equations of motion for the gluon field
$\textsl{v}_{\mu}$ field which arose from the \eq{EA91} action  are the following:
\beq\label{EA31}
\Le\,D_{\mu}\,G^{\mu \nu}\,\Ra_{a}\,=\,\partial_{\mu}\,G_{a}^{\mu \nu}\,+\,g\,f_{abc}\textsl{v}_{\mu}^{b}\,G^{c\,\mu \nu}\,=\,j_{a}^{+}\,\delta^{\nu +}\,+\,\,j_{a}^{-}\,\delta^{\nu -}
\eeq
We assume also, that the form of the currents in \eq{EA31} is not fixed yet, it will be demonstrated that it can be independently determined on the base of self-consistency request applied for the solutions of equations of motion, see \eq{Sol5} in the next section. Of course, this resulting current will be the same as the induced current \eq{EA9241} introduced in the previous Section.

  The light-cone gauge $\,\textsl{v}_{-}^{a}\,=\,0\,$ is applied further in the equations of motion and thereafter in the final effective action as well.
Considering \eq{EA31} we obtain the following equations for the different field components.  
\begin{enumerate}
\item
The	variation of the action with respect to  $\textsl{v}_{+}^{a}$  gives:
\beq\label{EA32}
-\,\partial_{i}\,\partial_{-}\,\textsl{v}^{i}_{a}\,-\,\partial^{+}\,\partial_{-}\,\textsl{v}_{a\,+}\,-\,
g\,f_{abc}\,\textsl{v}_{i}^{b}\,\Le\,\partial_{-}\,\textsl{v}^{i\,c}\,\Ra\,=\,j_{a}^{+}(\textsl{v}_{+})\,,
\eeq
or
\beq\label{EA321}
-\Le D_{i} \Le \partial_{-} \textsl{v}^{i}\,\Ra \Ra_{a}\,-\,\D^{2}_{-}\,\textsl{v}_{a\,+}\,=\,j_{a}^{+}(\textsl{v}_{+})\,.
\eeq

\item
  The variation on field $\textsl{v}_{-}^{a}$  provides
\beq\label{EA4}
\partial_{\mu}\,G^{\mu - }_{a}\,+\,g\,f_{abc}\,\textsl{v}^{b}_{\mu}\,G^{c \,\mu -}\,=\,
\,j^{-}_{a}(\textsl{v}_{-})\,,
\eeq
and, requesting the following current's property $j^{-}_{a}(\textsl{v}_{-}\,=\,0)\,=\,\,-\,\partial_{i}^{2}\,A_{a\,+}\,\,$, we obtain:
\beq\label{EA5}
\Le D^{-} G_{-+}\Ra_{a}\,+\,\Le D^{i}G_{i+}\Ra_{a}\,
=
\,-\,\partial_{i}^{2}\,A_{a\,+}\,.
\eeq
It can be written as following:
\beqar\label{EA6}
&\,&\D_{+}\,\D_{-}\,\textsl{v}_{a\,+} + g\,f_{abc}\,\textsl{v}^{b}_{+}\Le \partial_{-}\,\textsl{v}_{+}^{c}\Ra 
+  \partial^{i}
\Le
\partial_{i} \textsl{v}_{a\,+}  - \partial_{+} \textsl{v}_{a\,i} + g\,f_{abc}\textsl{v}^{b}_{i}\textsl{v}^{c}_{+}
\Ra + \\
 &\,& + g\,f_{abc}\,\textsl{v}^{b\,i}
\Le
\partial_{i} \textsl{v}_{+} ^{c} - \partial_{+} \textsl{v}_{i}^{c} + g\,f^{cdf}\textsl{v}_{d\,i}\textsl{v}_{f\,+}
\Ra\,=\,-\,\partial_{i}^{2}\,A_{a\,+}\,\nonumber\,,
\eeqar
and presented in one of the forms:
\beq\label{EA61}
\Le D_{+}[\textsl{v}_{+}]\Le\D_{-}\textsl{v}_{+}\Ra\Ra_{a}\,+\,
\Le\,D^{i}\Le\D_{i}\,\textsl{v}_{+}\Ra\Ra_{a}\,  - \,\D^{i}\Le D_{+} \textsl{v}_{i} \Ra_{a}- g\,f_{abc}\,\textsl{v}^{b\,i}
\Le D_{+} \textsl{v}_{i} \Ra^{c}\,=\,-\,\partial_{i}^{2}\,A_{a\,+}\,\,
\eeq
or
\beq\label{EA611}
\Le D_{+}[\textsl{v}_{+}]\Le\D_{-}\textsl{v}_{+}\Ra\Ra_{a}\,+\,
\Le\,D^{i}\Le\D_{i}\,\textsl{v}_{+}\Ra\Ra_{a}\,  - \,
\Le\,D^{i}[\textsl{v}_{i}]\,\Le\,D_{+}[\textsl{v}_{+}]\,\textsl{v}_{i}
\Ra\,\Ra_{a}\,=\,-\,
\partial_{i}^{2}\,A_{a\,+}\,,
\eeq
see Appendix C for more details. 
\item
  The variation on  field $\textsl{v}_{\bot}^{a}$ gives in turn:
\beq\label{EA8}
\Le D^{+} G_{+i}\Ra_{a}\,+\,\Le D^{-}G_{-i}\Ra_{a}\,+\,\Le D^{j}G_{ji}\Ra_{a}\,=\,0\,.
\eeq
Rewriting this equation as
\beq\label{EA9}
\partial_{-} \Le
\partial_{+} \textsl{v}_{a\,i} - \partial_{i} \textsl{v}_{a\,+} + g\,f_{abc}\textsl{v}^{b}_{+}\textsl{v}^{c}_{i}
\Ra +
\partial_{+}\partial_{-}\,\textsl{v}_{a\,i} +
g\,f_{abc}\,\textsl{v}^{b}_{+}\,\Le \partial_{-} \textsl{v}_{i}^{c}\Ra\,+\,\Le D^{j}G_{ji}\Ra_{a}\,=\,0\,.
\eeq
we finally obtain:
\beq\label{EA10}
 2\Le D_{+}\,\Le \D_{-} \textsl{v}_{i} \Ra\,\Ra_{a}\,-\,\Le\,D_{i}\,\Le \D_{-} \textsl{v}_{+} \Ra\,\Ra_{a}\,+\,\Le D^{j}G_{ji}\Ra_{a}\,=\,0\,.
\eeq
\end{enumerate}
The LO perturbative solutions of these equations are presented in the next section.

\section{The LO solution: from effective action approach to CGC expressions}

 For the action without the external source, at $j^{+}_{a}\,=\,0$, the trivial solution at the first perturbative order
can be easily obtained:
\beq\label{Sol1}
\textsl{v}_{\bot}\,=\,0\,,\,\,\textsl{v}_{+}\,=\,A_{+}\,,
\eeq
see \cite{LipatovEff}. 
At the presence of an external source the solution is changing and the following ansatz as 
solution of \eq{EA32}-\eq{EA10} can be introduced:
\beq\label{Sol11}
\textsl{v}_{-}\,=\,0\,,\,\,
\textsl{v}_{\bot}\,=\Lambda[g\,A_{+}]\,+\,g\,\Lambda_{1}[g\,A_{+}]\,+\,\cdots\,,\,\,
\textsl{v}_{+}\,=\,A_{+}\,+\,g\Phi[g\,A_{+},\textsl{v}_{\bot}]\,+\,\cdots\,,\,\, \D_{-}\,A_{+}\,=\,0\,,
\eeq
with still unknown form of the external current, which properties we will establish from the requests of self-consistency
of the solution. We note, that this ansatz is valid for the large values of reggeon field when $g\,A_{+}\,\sim\,1$  and in principal it can be
useful for the situations of non-symmetrical scattering processes such as DIS or scattering off nuclei. 

\begin{enumerate}
\item
 We begin from \eq{EA10} taking there $\D_{-} \textsl{v}_{+}\,=\,0$, see \eq{EA911} and \eq{Sol11}. We will see further that at LO $G_{i j}=0$, therefore the equation reads as  
\beq\label{Sol111}
\Le D_{+}\,\Le  \D_{-} \textsl{v}_{i} \Ra\,\Ra_{a}\,=\,0\,.
\eeq
It's solution has the following form
\beq\label{Sol2}
\D_{-}\,\textsl{v}_{i}^{b}\,=\,
U^{bc}\Le \textsl{v}_{+} \Ra\,
\D_{-}\rho_{c\,i}\Le x^{-} , x_{\bot}\Ra \,,
\eeq
with $U^{bc}$ as some function which form is determined by the request that it's  covariant derivative is equal to zero. The 
form of the function $\rho_{c\,i}\Le x^{-} , x_{\bot}\Ra$ is arbitrary in this case, it is proportional to the color charge density
in the CGC approach.
With LO precision, in the sense of \eq{Sol11} ansatz, we obtain:
\beq\label{Sol21}
\textsl{v}_{i}^{b}\,=\,U^{bc}\Le \textsl{v}_{+} \Ra\,\rho_{c i}\Le x^{-} , x_{\bot}\Ra \,.
\eeq
The form of $U^{a b}$ function in \eq{Sol21} can be written consistently with the  used in \cite{Kovner}:
\beq\label{Sol22}
\textsl{v}_{i}^{b}\,=\,tr\,[f^{b}\,U_{-\infty,\, x^{+}}\Le  \textsl{v}_{+} \Ra\,
f^{c}\,U_{x^{+},\, \infty}\,\Le  \textsl{v}_{+} \Ra\,]\,\rho_{c i}\Le x^{-} , x_{\bot}\Ra \,=\,
U^{b c}\rho_{c i}\Le x^{-} , x_{\bot}\Ra \,,
\eeq
where $U_{-\infty, x^{+}}\,=\,P\,e^{g \int_{-\infty}^{x^{+}}\,dx^{'+} \,\textsl{v}_{+}^{a}}$ and
which correspond to the form of the induced current in the effective action when $O$ operator is taken in the form of simple ordered exponential, see \eq{EA9241} and Appendixes A and B.

\item

The leading order solution of \eq{EA321} 
\beq\label{Sol3}
-\,D_{i}\,\partial_{-}\,\textsl{v}^{i}_{a}\,=\,j_{a}^{+}\,
\eeq 
will determine the form of unknown function $\rho_{c i}\Le x^{-} , x_{\bot}\Ra \,$ 
in terms of the given external current\footnote{In the given framework  we can take l.h.s. of \eq{Sol3} as definition of the external current.} 
$j_{a}^{+}$, assumed to be unknown for the moment. Taking \eq{Sol3} to LO, we obtain:
\beq\label{Sol33}
-\,\D_{i}\,\partial_{-}\,\textsl{v}^{i}_{a}\,=\,j_{a}^{+}\,.
\eeq 
The current in the r.h.s. of \eq{Sol33} we write in the form self-consistent with \eq{Sol22}:
\beq\label{Sol5}
 j_{a}^{+}\,=\,-\,U^{ab}\,\Le \textsl{v}_{+} \Ra\,\tilde{J}^{+}_{b}\Le x^{-} , x_{\bot}\Ra\,,
\eeq
we see, that this condition of self-consistency  dictates the same form of the current as induced current introduced in \eq{EA9111}, see \eq{EA9241}.
Now, we have to the first perturbative order :
\beq\label{Sol31}
\partial_{i}\,\partial_{-}\,\rho^{i}_{a}\,=\,\tilde{J}^{+}_{a}\Le x^{-} , x_{\bot}\Ra\,,
\eeq 
which is the same equation as in \cite{Kovner}.
The \eq{Sol31}  has no simple solution and, following to \cite{Kovner}, we assume the following structure of this term :
\beq\label{Sol32}
\,\tilde{J}^{+}_{a}\Le x^{-} , x_{\bot}\Ra\,=\,\D_{i}\,\D_{-}\,\tilde{j}_{a}^{i}\Le x^{-} , x_{\bot}\Ra\,,
\eeq
that gives
\beq\label{Sol6}
\rho_{c}^{i}\Le x^{-} , x_{\bot}\Ra\,=\,\tilde{j}_{c }^{i}\Le x^{-} , x_{\bot}\Ra\,.
\eeq
In CGC approach the another assumption  is done, namely it is assumed in \cite{Venug,Kovner} that
\beq\label{Sol511}
\D_{-}\,\tilde{j}_{a}^{i}\Le x^{-} , x_{\bot}\Ra\,=\,\delta(x^{-})\,\tilde{\rho}_{c }^{i}\,\Le \,x_{\bot}\,\Ra\,,
\eeq
with some known $\tilde{\rho}_{c }^{i}\,\Le \,x_{\bot}\,\Ra\,$ functions.
We note, that this assumption providing the factorization between $x^{-}$ and $x_{\bot}$  coordinates in the LO leads 
to the difficulties in the NLO solution, it is seen already from \eq{EA321}. Still, accepting this assumption, we obtain:
\beq\label{Sol52}
\tilde{j}_{c }^{i}\Le x^{-} , x_{\bot}\Ra\,=\,\theta( x^{-})\,\tilde{\rho}_{c }^{i}\,\Le \,x_{\bot}\,\Ra\,,
\eeq
that provides
\beq\label{Sol521}
\rho_{c}^{i}\Le x^{-} , x_{\bot}\Ra\,=\,\theta(x^{-})\,\rho_{c}^{i}\Le x_{\bot}\Ra\,=\,\theta(x^{-})\,\tilde{\rho}_{c }^{i}\,\Le \,x_{\bot}\,\Ra\,
\eeq
and correspondingly
\beq\label{Sol53}
\textsl{v}_{i}^{b}\Le  x^{+} , x_{\bot} \Ra \,=\,
\theta(x^{-})\,U^{bc}\,\Le \textsl{v}_{+} \Ra\,\rho_{c i}\Le x_{\bot}\Ra \,,
\eeq
where the first term of perturbative expansion of \eq{Sol53} is in full agreement with the LO CGC result of \cite{Kovner}.

\item
 For the last equation of motion, \eq{EA6}, we have 
at the first order
\beq\label{Sol71}
 \D_{i}\,\D^{i}\, \textsl{v}_{a\,+}\,=\,-\,\D_{i}^{2} A_{a\,+}\,,
\eeq
that gives  
\beq\label{Sol92}
\textsl{v}_{a\, +}\,=\,\,A_{a\,+}\,,
\eeq
in correspondence with \eq{Sol11}, see \cite{LipatovEff}.
In turn, it provides the first order solution for the field $\textsl{v}_{i}$: 
\beq\label{Sol10}
\textsl{v}_{i}^{b}\Le  x^{+} , x_{\bot} \Ra \,=\,U^{bc}\,\Le A_{+} \Ra\,\rho_{c i}\Le x^{-}\,, x_{\bot}\Ra \,.
\eeq
We underlie, that this solution consists all orders of $g$ through the ordered exponential $\,U^{bc}\,$, that is the novel result of our calculations.
\end{enumerate}

 In order to relate obtained LO solutions \eq{Sol92}-\eq{Sol10} with CGC results, 
we note, that the results of \cite{Kovner} for the  classical gluon 
field are reproduced taking $A_{+}\,=\,0$ in these expressions.

\section{LO structure of the effective action}

 Calculations in the previous section we can consider as a formulation of RFT calculus based on the effective action approach.
In this case, basing on the \eq{ApB4} for the current, we consider solution of equations of motion
as solution for the classical gluon field in the presence of $A_{-}$ source. This transition can be done by the following substitution:
\beq\label{RF1}
\tilde{J}_{a}^{+}\,\rightarrow\,-\,\D_{\bot}^{2}\,A_{a}^{+}\,,
\eeq 
see \eq{ApB4} again and discussions in \cite{Kovner} and \cite{Hatta1}.
Therefore, instead of \eq{Sol31}, we obtain:
\beq\label{RF2}
\D_{i}\,\D_{-}\,\rho_{a}^{i}\,=\,-\,\frac{1}{N}\,\D_{\bot}^{2}\,A_{a}^{+}\,,
\eeq
or
\beq\label{RF3}
\rho_{a}^{i}\,=\,\frac{1}{N}\,\D_{-}^{-1}\,\Le\D^{i}\,A_{-}^{a}\Ra\,,
\eeq
where the condition $G_{i j}\,=\,0$ is provided at LO approximation. Correspondingly, all results of the previous section can be 
rewritten and we obtain for \eq{Sol22}:
\beq\label{RF4}
\textsl{v}_{i}^{a}\,=\,\frac{1}{N}\,U^{a b}\,\Le \textsl{v}_{+} \Ra\,
\Le\,\D_{-}^{-1}\,\Le\D_{i}\,A_{-}^{b}\Ra\,\Ra\,.
\eeq
The \eq{Sol92} solution for field $\textsl{v}_{+}$  remains unchanged under substitution \eq{RF3} as well as form of the NLO solution
of Appendix C.

 Inserting obtained classical gluon fields solutions in the \eq{EA92} action, we will obtain a 
action which will  depend only on the reggeon fields, see \cite{LipatovEff}, determining the LO RFT  action of the approach.
Formally, due to the presence of ordered path exponential in the solutions, the action will includes all order 
perturbative terms which can be important for large $v_{+}\,\approx\,A_{+}$ in the processes where some large color charge is created. The expansion of these exponential must be supplemented by  solution of equations of motion to corresponding orders, otherwise only part of the usual perturbative corrections will be accounted.
In general, the following expansion for  the action exists:
\beq\label{RFA1}
S_{eff}\,=\,-\,\int\,d^{4}\,x\,\Le\,s_{1}[g,\,A_{+},\,A_{-}]\,+\,g\,s_{2}[g,\,A_{+},\,A_{-}]\,+\,\cdots\,\Ra\,,
\eeq
where additional dependence on the coupling constants in the different terms of the Lagrangian is arising through the ordered exponentials in the classical solutions for the gluon field
accordingly to the ansatz of \eq{Sol11}.

  In order to calculate
this action we need to know the components of the field strength tensor, with LO precision we have:
\beq\label{RFA2} 
G_{+\,-}^{a}\,=\,0\,,\,\,G_{i\,+}^{a}\,=\,\D_{i}\,A_{+}^{a}\,,\,\,
G_{i\,-}^{a}\,=\,-\,\D_{-}\,\textsl{v}_{i}^{a}\,,\,\,G_{i\,j}\,=\,0\,,
\eeq 
that gives
\beq\label{RFA3}
\frac{1}{4}\,G_{\mu\,\nu}\,G^{\mu\,\nu}\,=\,-\,G_{i\,-}\,G_{i\,+}\,=\,\Le\,\D_{-}\,\textsl{v}_{i}^{a}\,\Ra\,.
\Le\,\D_{i}\,A_{+}^{a}\,\Ra\,.
\eeq
Therefore, for the \eq{EA91} effective action we obtain to LO:
\beq\label{RFA4}
S_{eff}=\,-\,\int d^{4}x \Le\,\Le\D_{i} A_{+}^{a}\Ra\,
U^{a\,b}(A_{+})\,\Le\D_{-}\rho_{i\,b}(x_{\bot})\Ra\,+\,\frac{1}{N}\,
A_{+}^{a} \,\Le\,O^{a\,b}(A_{+})\,+\,N\,\delta^{a\,b}\,\Ra\,\Le \D_{\bot}^{2}\,A^{b}_{-}\Ra\,\Ra\,,
\eeq
with $O^{a\,b}(A_{+})\,=\,Tr\left[f^{a}\,O(A_{+})\,f^{b}\,\right]$ in adjoint representaion. Using \eq{RF3} we rewrite this expression as
\beq\label{RFA5}
S_{eff}=\,-\,\frac{1}{N}\,\int d^{4}x \Le\,\Le\D_{i} A_{+}^{a}\Ra\,
U^{a\,b}(A_{+})\,\Le \D_{i}\,A_{-}^{b} \Ra\,+\,
A_{+}^{a} \,\Le\,O^{a\,b}(A_{+})\,+\,N\,\delta^{a\,b}\,\Ra\,\Le \D_{\bot}^{2}\,A^{b}_{-}\Ra\,
\Ra\,.
\eeq
We obtained, that due the ordered exponential in the action, there are some additional corrections 
which were not considered in 
\cite{LipatovEff}. However, obtained in expression \eq{RFA5} corrections are not complete,
we also need the higher order solutions of equations of motion.
Thus, using  results of Appendix C, we can calculate full tree NLO   corrections to the 
 reggeon-reggeon transition vertices as well, but  we postpone this task for the following publications.

\section{Effective action for  reggeized gluons and CGC approach}

 In the CGC approach, auxiliary soft and semi-hard gluon fields are added to classical gluon field. Integrating the semi-hard  fields out, some effective action is obtained and
relations similar to \eq{Sol92}, \eq{Sol10}  are used there as an initial condition for the further small-x evolution of the gluon density operator. 
Therefore, it will be  constructive to determine the counterpart to the reggeon field in the CGC approach. 
From \cite{Kovner} we know, that the classical gluon field $\textsl{v}^{a}_{+\,cl}$ is zero there and only the 
fluctuations of this field are considered
\beq\label{CGC1}
\textsl{v}_{+\,cl}\,=\,\delta v_{+}\,+\,a_{+}\,,
\eeq
which are ordered in the longitudinal momenta. There are semi-hard and soft fluctuations of the field, $\delta v_{+}\,$ and  $a_{+}$ correspondingly, see details in \cite{Kovner}.
By direct comparison of this field's representation with \eq{Sol92}, we see that the $\delta v_{+}\,$ fluctuation in the CGC is precisely
$A_{+}$ reggeon field in the effective action, see also kinematic properties of the field in \cite{LipatovEff} and 
\cite{Kovner}. The difference between two approaches is that whereas the reggeon fields are present in the effective action initially as some parameter of the problem,
which must be considered separately after all in the path integral as independent fields, 
in the CGC approach the reggeon like field $\delta v_{+}$ appears as fluctuation around $\textsl{v}^{a}_{+\,cl}\,=\,0$ classical solution and integrated out. As we obtained above,
taking $A_{+}\,=\,0$ in the \eq{Sol22} and \eq{Sol92} we will reproduce the CGC answers for the LO  classical fields configuration. 
The counterpart of  gluon density operator (source) in the CGC calculations is $A_{-}$, the second reggeon field, see \eq{RF2}-\eq{RF3}, that allows to relate results of the approaches. 
We also note, that the classical gluon field \eq{Sol21} consists terms to all order of coupling constant, that differ it from the classical solutions considered in the CGC approach.

  There is an important point that requires further clarification. Expansion of the action in terms of background field 
(reggeon field or $\delta\,v_{+}\,$ fluctuation) requires solution of equations of motion for gluon fields at the same order
of background field. Namely, taking non-zero $\delta v_{+}\,$ we will obtain that the NLO solution for the gluon field will  depend on 
$\delta v_{+}\,$  already, see Appendix C. Thereby, expanding the Lagrangian in terms of background fields,
we have to account the same order terms which arise also from the square of field strength tensor and from the currents in the action. 
Indeed, there is NLO solution of equations of motion, see Appendix C, which after the insertion into the Lagrangian will 
reproduce the same order terms as in the expansion of the effective currents in respect to the soft fluctuations, these corrections are absent in CGC approach,
see also  \cite{Hatta2}, \cite{Hatta1} where the NLO correction to the CGC framework results were discussed.

  
	We also note, that both approaches give the similar equations at the level of equations of motion, when the simplest form of $O$ and $O^{T}$ operators are used, see Appendices B and C.
This is related to the fact that the additional term in the CGC action can be considered as resulting from the integration on $x_{-}$ coordinate of the current term in the effective action for reggeized gluons, see Iancu et al. in \cite{Kovner} and \cite{Hatta1} for the relevant discussions.

\section{Conclusion}

 In this paper we consider application of the effective action approach for reggeized gluons to the calculation of a classical gluon field produced by relativistic color charge. 
We demonstrate, that effective action for reggeized gluons can be obtained from QCD action, when both reggeon fields are introduced as non-zero LO solutions 
for the classical longitudinal gluon fields and as sources of each other. 
The form of effective currents in the action in this case,
can be obtained from the request of self-consistency of classical equations of motion.
We also obtained \eq{Sol92}, \eq{Sol10}, \eq{App811} and \eq{App13} expressions 
for the classical gluon fields calculated in the effective action formalism, which consist all order terms in respect to the coupling constant,
this is a main result of the paper. In these calculations the results  for the classical gluon fields of CGC approach  can be reproduced  in the limit of zero reggeon fields, 
see discussion in the previous Section. 

 In the framework of the effective action, the performed calculations can be considered as 
solution of classical equations of motion with the reggeons fields introduced as LO classical solutions of longitudinal gluon fields.
Considering   field $A_{-}$ in the as an external source, see \eq{RF2}-\eq{RF3}, the same problem can be understood as calculation of some effective particle-reggeon-particle vertex, see  expression \eq{EA91}.
In this case, introducing fluctuations around the classical solution and integrating them out, an 
one loop correction to this vertex can be obtained. 
In general, we can determine the effective action for reggeized gluons in terms of reggeon fields only.
Inserting found classical solutions solutions in the action we will obtain some effective action for the reggeon fields $A_{+}$ and $A_{-}$ in the following form:
\beq\label{EK2}
\Gamma\,=\,\sum_{n,m\,=\,0}\,\Le\,A_{+}^{a_{1}}\,\cdots\,A_{+}^{a_{n}}\,K^{a_1\,\cdots\,a_{n}}_{b_1\,\cdots\,b_{m}}\,A_{-}^{b_{1}}\,\cdots\,A_{-}^{b_{m}}\,\Ra\,,
\eeq
that will allow to calculate different reggeon-reggeon transition vertices $\,K^{a_1\,\cdots\,a_{n}}_{b_1\,\cdots\,b_{m}}\,$ 
responsible for the unitarization of the scattering amplitudes at high energies, which also can be used
in calculations of the amplitudes of various processes with  multi-Regge and quasi-multi-Regge kinematics.

  In general, it will be interesting  to investigate possible relations of the considered framework with results of 
\cite{Hatta2},\cite{Bal}  and
we hope that
further work in the proposed direction will allow to establish useful correspondences between the different small $x$ approaches and calculate high order corrections 
 to the amplitudes of high-energy scattering.

\section{Acknowledgments}

 The authors are thankful to E.Levin and S.Pozdnyakov for helpful discussions concerning the subject of the paper and an especial 
thank to Jochen Bartels for his hospitality at Hamburg, where this project was initiated. L.L. acknowledges partial support of the Russian Scientific Fund project 14-2200281 and
would like to thank the State University of St. Petersburg
for the grant
SPSU 11.38.223.2015 and the grant RFBI 16-02-01143 for support.


\newpage
\newpage
\section*{Appendix A: Induced current in the effective action}
\renewcommand{\theequation}{A.\arabic{equation}}
\setcounter{equation}{0}

 In this Appendix we consider an $j^{+}$ component of the induced current which can be obtained
by variation of the current term in the effective action \eq{EA92}:
\beq\label{ApB1}
j^{+}_{ind}(\textsl{v}_{+})\,=\,j_{-}^{ind}(\textsl{v}_{+})\,=\,
\frac{1}{N}\,O(\textsl{v}_{+})\,
\Le\,\D_{i}^{2}\,A^{+}\,\Ra\,O^{T}(\textsl{v}_{+})\,.
\eeq
The operators $O$ and $O^{T}$ are introduced in \cite{LipatovEff}: 
\beq\label{ApB6}
O\, =\,\D_{+}\,\Le\,D^{-1}_{+}\,\Ra\,\,;\,\,
 O^{T} \, =\,\Le\,D^{-1}_{+}\,\Ra\,\overleftarrow{\D}_{+}\,,
\eeq
and have the following properties:
\beq\label{ApB7}
\D_{+}\,O\,=\,g\,\textsl{v}_{+}\,O\,\,,\,\,O^{T}\,\overleftarrow{\D}_{+}\,=\,-\,g\,O^{T}\,\textsl{v}_{+}\,,
\eeq
see Appendix B further.
The \eq{ApB1} form of the current is general,
the particular representations of the current in terms of P-exponentials, in turn, depend
on the representations of $O$  operator, or, more precisely on the
representation of the $\D_{+}^{-1}$ operator.
If we take the following simplest representation
\beq\label{ApB8}
\frac{1}{\D_{+}}\,f(x^{+})\,=\,\int^{x^{+}}_{-\infty}\,dx^{'+}\,f(x^{'+})\,,
\eeq
we obtain for these operators:
\beq\label{ApB2}
O \, =\, P\,e^{\imath g \int_{-\infty}^{x^{+}}\,dx^{'+} \,\textsl{v}_{+}^{a}\,T_{a} }\,
\eeq
and
\beq\label{ApB3}
O^{T} \, =\, P\,e^{\imath g \int^{\infty}_{x^{+}}\,dx^{'+} \,\textsl{v}_{+}^{a}\,T_{a} }\,.
\eeq

  The variation of interaction term in the action can be calculated with the help of the formulas from Appendix B
and results by induced current from \eq{ApB1}:
\beq\label{ApB11}
\delta\,\Le\,A_{+}(\textsl{v}_{+})\,j^{+}_{reg}\Ra\,=\,-\,\imath\,\Le\delta\textsl{v}_{+}^{a}\Ra\,tr[T_{a}\,j^{+}_{ind}(\textsl{v}_{+})]\,=\,
\Le\delta\textsl{v}_{+}^{a}\Ra\,j_{a}^{+}(\textsl{v}_{+})\,=\,-\,\frac{1}{N}\,
\Le\delta\textsl{v}_{+}^{a}\Ra\,tr\,[\,T_{a}\,O\,T_{b}\,O^{T}\,]\,\Le\,\D_{i}^{2}\,A^{+}_{b}\,\Ra\,,
\eeq 
with $v_{+}\,=\,\imath\,T^{a}\,v_{+}^{a}$ representation of the gluon field used.
In the case of adjoint representation\footnote{The general form of the 
current does not depend on the representation, in our particular case we take 
$\Le\,T_{a}\,\Ra_{b\,c}\,=\,-\,\imath\,f_{a\,b\,c}$
in the current, representing only "external" indexes in the expression.} we will obtain:
\beq\label{ApB4}
\delta\Le A_{+}(\textsl{v}_{+})\,j^{+}_{reg}\Ra\,=\,\Le\delta \textsl{v}_{+}^{a}\Ra\,j_{a}^{+}(\textsl{v}_{+})\,=\,
\frac{1}{N}\,
\Le \delta\textsl{v}_{+}^{a}\Ra\,tr\,[\,f_{a}\,O\,f_{b}\,O^{T}\,]\,\Le\,\D_{i}^{2}\,A^{+}_{b}\,\Ra\,=\,\frac{1}{N}\,
\Le \delta\textsl{v}_{+}^{a}\Ra\,U^{a\, b}\,\Le\,\D_{i}^{2}\,A^{+}_{b}\,\Ra\,,
\eeq 
that provides
\beq\label{ApB44}
j_{a}^{+}(\textsl{v}_{+}\,=\,0)\,=\,-\,\D_{i}^{2}\,A^{+}_{a}\,.
\eeq
The $U^{a\, b}\,$ exponential in \eq{ApB4} is the same as used in
CGC approach of  \cite{Kovner}, see \eq{Sol22}.
The \eq{ApB2}-\eq{ApB3} forms of the operators can be modified in order to provide the action's unitarity at
$x\rightarrow\,\pm\infty$.
For that, the \eq{ApB8} operator can be modified as\footnote{We use the Kogut-Soper convention for the metric tensor, in the Lepage-Brodsky convention there is $1/4$ coefficient in the front of the following expression.}:
\beq\label{ApB9}
\frac{1}{\D_{+}}\,f(x^{+})\,=\,\frac{1}{2}\int\,dx^{'+}\,\epsilon\,
\Le x^{+}\,-\,x^{'+} \Ra\,f(x^{'+})\,,
\eeq
where $\epsilon\,\Le x^{+}\,-\,x^{'+} \Ra\,$ is a sign function,
that corresponds to the different from \eq{ApB8} definition of the integral operator $\D_{+}^{-1}$, where the 
regularization of the corresponding $1/k_{+}$ pole
in momentum space must be understood as principal value prescription, see details in \cite{LipatovEff}. In this case, more complicated expressions for the
operators will be obtained, see also Appendix B below.

\newpage
\section*{Appendix B: Representation and properties of  operators $O$ and $O^{T}$}
\renewcommand{\theequation}{B.\arabic{equation}}
\setcounter{equation}{0}

For the arbitrary representation of gauge field $v_{+}\,=\,\imath\,T^{a}\,v_{+}^{a}$ with
$D_{+}\,=\,\D_{+}\,-\,g\,v_{+}$, we can consider
the following representation of $O$ and $O^{T}$ operators
\footnote{Due the light cone gauge we consider here only $O(x^{+})$ operators. 
The construction of the representation of the $O(x^{-})$ operators can be done similarly. We also note, that the integration is assumed for 
repeating indexes in expressions below if it is not noted otherwise. }:
\beq\label{A11}
O_{x}\,=\,\delta^{a\, b}\,+\,g\,\int\,d^{4}y\,G_{x y}^{+\,a\, a_{1}}\, \Le v_{+}(y)\Ra_{a_1\,b} \, =\,
\,1\,+\,g\,G_{x y}^{+}\, v_{+ y}\,
\eeq
and correspondingly
\beq\label{A12}
O^{T}_{x}\,=\,\,1\,+\,g\,v_{+ y}\,G_{y x}^{+}\,,
\eeq
which is redefinition of the operator expansions used in \cite{LipatovEff} in terms of Green's function instead 
integral operators, see Appendex B above.
The Green's function in above equations we understand as Green's function of the $D_{+}$ operator
and express it in the perturbative sense as :
\beq\label{A13}
G_{x y}^{+}\,=\,G_{x y}^{+\,0}\,+\,g\,G_{x z}^{+\,0}\, v_{+ z}\,G_{z y}^{+}\,
\eeq
and
\beq\label{A14}
G_{y x}^{+}\,=\,G_{y x}^{+\,0}\,+\,g\,G_{y z}^{+}\, v_{+ z}\,G_{z x}^{+\,0}\,,
\eeq
with the bare propagators defined as (there is no integration on index $x$ in expressions)
\beq\label{A15}
\D_{+ x}\,\,G_{x y}^{+\,0}\,=\,\delta_{x\,y}\,,\,\,\,G_{y x}^{+\,0}\,\overleftarrow{\D}_{+ x}\,=\,-\delta_{x\,y}\,.
\eeq
The following properties of the operators now can be derived:
\begin{enumerate}
\item
\beqar\label{A161}
\delta\,G^{+}_{x y}& = & g\,G_{x z}^{+\,0}\,\Le \delta v_{+ z} \Ra\,G_{z y}^{+}+
\,G_{x z}^{+\,0}\, v_{+ z}\,\delta G_{z y}^{+}=
g\,G_{x z}^{+\,0}\,\Le \delta v_{+ z} \Ra\,G_{z y}^{+}+
\,G_{x z}^{+\,0}\, v_{+ z}\,\Le \delta G_{z p}^{+} \Ra\,D_{+ p}\,G^{+}_{p y}=\nonumber \\
&=&
g \Le
G_{x z}^{+\,0}\,\Le \delta v_{+ z} \Ra\,G_{z y}^{+}-
G_{x z}^{+\,0}\, v_{+ z}\,G_{z p}^{+}\,\Le \delta D_{+ p}\Ra\,G^{+}_{p y}\Ra=
g \Le
G_{x p}^{+\,0}\,+\,G_{x z}^{+\,0}\, v_{+ z}\,G_{z p}^{+}\Ra\,\delta v_{+ p} \,G^{+}_{p y}=\nonumber \\
&=&\,
\,g\,G^{+}_{x p}\,\delta v_{+ p} \,\,G^{+}_{p y}\,;
\eeqar
\item
\beq\label{A16}
\delta\,O_{x}\, = \,g\,G^{+}_{x y}\,\Le \delta v_{+ y} \Ra\,+g\,\Le \delta G^{+}_{x y}\Ra\,v_{+ y}\,=
\,g\,G^{+}_{x p}\,\delta v_{+ p}\,\Le 1\, +\,g \,G^{+}_{p y}\,v_{+ y}\,\Ra\,=\,
g\,G^{+}_{x p}\,\delta v_{+ p}\,O_{p}\,;
\eeq
\item
\beq\label{A17}
\D_{+ x}\,\delta\,O_{x}\,=\,g\,\Le \D_{+ x}\,G^{+}_{x p} \Ra\,\delta v_{+ p}\,O_{p}\,=\,
g\,\Le 1\,+\,g\,\,v_{+ x}\,G^{+}_{x p}\,\Ra\,\delta v_{+ p}\,O_{p}\,=\,
g\,O^{T}_{x}\,\delta v_{+ x}\,O_{x}\,;
\eeq
\item
\beq\label{A18}
\D_{+ x}\,O_{x}\,=\,g\,\Le \D_{+ x}\,G^{+}_{x y} \Ra\,v_{+ y}\,=\,
g\,v_{+ x}\,\Le 1\,+\,g\,G_{x y}^{+}\,v_{+ y}\,\Ra\,=\,
g\,v_{+ x}\,O_{x}\,;
\eeq
\item
\beq\label{A19}
O_{x}^{T}\,\overleftarrow{\D}_{+ x}\,=\,g\,v_{+ y}\,\Le G^{+}_{y x}\,\overleftarrow{\D}_{+ x} \Ra\,=\,
-\,g\,\Le 1\,+\,v_{+ y}\,\,G^{+}_{y x}\,\Ra\,v_{+ x}\,=\,-g\,O^{T}_{x}\,v_{+ x}\,.
\eeq
\end{enumerate}
We see, that the operator $O$ and $O^{T}$ have the properties of ordered exponents. For example, choosing bare propagators as
\beq\label{A110}
\,G_{x y}^{+\,0}\,=\,\theta(x^{+}\,-\,y^{+})\,\delta^{3}_{x y}\,,\,\,\,
\,G_{y x}^{+\,0}\,=\,\theta(y^{+}\,-\,x^{+})\,\delta^{3}_{x y}\,,\,\,\,
\eeq
we immediately reproduce:
\beq\label{A111}
O_{x}\,=\,P\, e^{g\int_{-\infty}^{x^{+}}\,dx^{'+}\, v_{+}(x^{'+})} \,,\,\,\,
O^{T}_{x}\,=\,P\, e^{g\int_{x^{+}}^{\infty}\,dx^{'+}\, v_{+}(x^{'+})} \,.
\eeq
The form of the bare propagator
$
\,G_{x y}^{+\,0}\,=\,\frac{1}{2}\,\left[\,\theta(x^{+}\,-\,y^{+})\,-\,\theta(y^{+}\,-\,x^{+})\,\right]\,\delta^{3}_{x y}\,
$ 
which correspond to the \eq{ApB9} integral operator
will lead to the
more complicated representations of $O$ and $O^{T}$ operators, see in \cite{LipatovEff}.

  Now we consider a variation of the action's full current :
\beq\label{A112}
\delta\, tr[v_{+ x}\,O_{x}\,\D_{i}^{2}\,A^{+}]=
\frac{1}{g}\,\delta\, tr[\Le \D_{+ x}\,O_{x} \Ra \D_{i}^{2}\,A^{+}]=\frac{1}{g}\, tr[
\Le\D_{+ x} \delta \,O_{x} \Ra \D_{i}^{2}\,A^{+}] = tr[
O^{T}_{x}\,\delta v_{+ x}\,O_{x}\Le \D_{i}^{2}\,A^{+}\Ra]\,,
\eeq 
which can be rewritten in the familiar form used in the paper:
\beq\label{A1121}
\delta\,\Le v_{+}\,J^{+} \Ra\,=\delta\,tr[\, \Le\, v_{+ x}\,O_{x}\,\D_{i}^{2}\,A^{+}\,\Ra\,]\,=\,-\,
\delta v_{+}^{a}\,tr[\,T_{a}\,O\,T_{b}\,O^{T}\,]\,\Le \D_{i}^{2} A^{+}_{b}\Ra\,.
\eeq
We also note, that with the help of \eq{A11} representation of the $O$ operator
the full action's  current can we written as follows
\beq\label{A113}
 tr[\Le v_{+ x}\,O_{x}\,-\,A_{+}\Ra\,\D_{i}^{2}\,A^{+}\,]\,=\,
tr[\Le v_{+}\,-\,A_{+} + v_{+ x}\,G^{+}_{x y}\,v_{+ y}\,\Ra\,\Le \D_{i}^{2} A^{+}\Ra]\,.
\eeq 

\newpage
\section*{Appendix C: NLO solution of equations of motion}

\renewcommand{\theequation}{C.\arabic{equation}}
\setcounter{equation}{0}

 In this Appendix we derive expressions for the 
next order solution of the equations of motion. We write the longitudinal field of interests as
\beq\label{App1}
\textsl{v}_{+}^{a}\,=
A_{+}^{a}\,+\,g\,\textsl{v}_{+1}^{a}(x_{\bot}, x^{-}, x^{+})\,,
\eeq 
and the transverse field in the next order approximation as:
\beq\label{App2}
\textsl{v}_{i}^{a}\,
=\,\textsl{v}_{i\,0}^{a}\,+\,
g\,\textsl{v}_{i\,1}^{a}(x_{\bot}, x^{-}, x^{+})\,+\,
g\,\tilde{\textsl{v}}_{i\,1}^{a}(x_{\bot}, x^{-}, x^{+})\,
\eeq
with the following constraint
\beq\label{App211}
\D^{i}\,\textsl{v}_{i\,1}^{a}\,=\,0\,,
\eeq
where
\beq\label{App21}
\textsl{v}_{i\,0}^{a}\,=\,\rho^{b}_{i}(x_{\bot},x^{-})\,U^{a\,b}(A_{+})\,.
\eeq

\begin{enumerate}
\item 
Let's consider again  the equations of motion and will begin from \eq{EA321}:  
\beq\label{App3} 
-\Le D_{i} \Le \partial_{-} \textsl{v}^{i}\,\Ra \Ra_{a}\,-\,\D^{2}_{-}\,\textsl{v}_{a\,+}\,=\,j_{a}^{+}(A_{+})\, 
\eeq 
which at requested order has the following form:
\beq\label{App4}
-\,g\,\D^{i}\D_{-}\,\tldv_{i\, 1}^{a}\,-\,\Le\D^{i} U^{a b}\Ra\,\Le\D_{-}\rho_{i}^{b}\Ra\,-\,g\,f_{a b c}\,
\Le U^{b\,b^{'}} \,\rho^{i b^{'}}\Ra\,\Le \,U^{c\,c^{'}}\,\Le\D_{-}\rho^{c^{'}}_{i}\Ra\Ra\,-\,g\,
\D_{-}^{2}\,\textsl{v}_{+1}^{a}\,=\,0\,.
\eeq
Denoting
\beq\label{App15}
g\,j^{+}_{a\,1}\,=\,g\,f_{a b c}\,
\Le U^{b\,b^{'}} \,\rho^{i b^{'}}\Ra\,\Le \,U^{c\,c^{'}}\,\Le\D_{-}\rho^{c^{'}}_{i}\Ra\Ra\,,
\eeq
we obtain:
\beq\label{App5}
\textsl{v}_{+1}^{a}\,=\,-\,\D^{i}\D_{-}^{-1}\,\tldv_{i\, 1}^{a}\,-\,\frac{1}{g}\,\Le\D^{i} U^{a b}\Ra\,
\Le\D_{-}^{-1}\rho_{i}^{b}\Ra\,-\,\Le \D_{-}^{-2}\,j^{+}_{a\,1}\Ra\,,
\eeq 
or
\beq\label{App6}
\,\D^{i}\,\tldv_{i\, 1}^{a}\,=\,-\,\D_{-}\,\textsl{v}_{+1}^{a}\,-\,\frac{1}{g}\,\Le\D^{i} U^{a b}\Ra\,\rho_{i}^{b}\,
-\,\Le \D_{-}^{-1}\,j^{+}_{a\,1}\Ra\,.
\eeq

\item
 Now we consider  equation of motion \eq{EA611}
\beq\label{App61}
\Le D_{+}[\textsl{v}_{+}]\Le\D_{-}\textsl{v}_{+}\Ra\Ra_{a}\,+\,
\Le\,D^{i}\Le\D_{i}\,\textsl{v}_{+}\Ra\Ra_{a}\,  - \,
\Le\,D^{i}[\textsl{v}_{i}]\,\Le\,D_{+}[\textsl{v}_{+}]\,\textsl{v}_{i}
\Ra\,\Ra_{a}\,=\,-\,
\partial_{i}^{2}\,A_{a\,+}\,
\eeq
which at NLO reads as
\beq\label{App7}
\D_{+}\,\D_{-}\,\textsl{v}_{+ 1}^{a}\,+\,\D_{i}\D^{i}\,\textsl{v}_{+ 1}^{a}\,-\,
\D_{+}\,\D^{i}\,\tldv_{i 1}^{a}\,+\,f_{a b c}\,U^{b\,b^{'}}\,
\rho^{i}_{b^{'}}\,\D_{i}\,A_{+}^{c}\,=\,0\,.
\eeq 
or
\beq\label{App71}
\D_{+}\,\D_{-}\,\textsl{v}_{+ 1}^{a}\,+\,\D_{i}\D^{i}\,\textsl{v}_{+ 1}^{a}\,-\,
\D_{+}\,\D^{i}\,\tldv_{i 1}^{a}\,+\,\frac{1}{g}\,\Le\D_{+}\D^{i} U^{a b}\Ra\,\rho_{i}^{b}\,=\,0\,.
\eeq
Inserting \eq{App6} into \eq{App7} one obtains
\beq\label{App8}
\Le\,2\,\D_{+}\,\D_{-}\,+\,\D_{i}\D^{i}\,\Ra\,\textsl{v}_{+ 1}^{a}\,=\,
\Box\,\textsl{v}_{+ 1}^{a}\,=\,-\,\frac{1}{g}\,\Le\D_{+}\D^{i} U^{a b}\Ra\,\rho_{i}^{b}\,
-\,f_{a b c}\,U^{b\,b^{'}}\,
\rho^{i}_{b^{'}}\,\D_{i}\,A_{+}^{c}\,-\,\Le \D_{+}\D_{-}^{-1}\,j^{+}_{a\,1}\Ra\,,
\eeq
that gives
\beq\label{App81}
\textsl{v}_{+ 1}^{a}\,=\,-\,\frac{2}{g}\,\Box^{-1}\,\Le\,\Le\D_{+}\D^{i} U^{a b}\Ra\,\rho_{i}^{b}\,\Ra\,-\,
\Box^{-1}\,\Le \D_{+}\D_{-}^{-1}\,j^{+}_{a\,1}\Ra\,.
\eeq
Taking into account that the last term in \eq{App81} is of  order $g^2$, finally for this field we have:
\beq\label{App811}
\textsl{v}_{+ 1}^{a}\,=\,-\,\frac{2}{g}\,\Box^{-1}\,\Le\,\Le\D_{+}\D^{i} U^{a b}\Ra\,\rho_{i}^{b}\,\Ra\,.
\eeq
In turn, inserting \eq{App5} into \eq{App71}, one obtains
\beq\label{App9}
2\D_{+}\,\Le \D^{j}\,\tldv_{j\,1}^{a}\Ra\,+\,\D_{j}\D^{j}\,\Le \D^{-1}_{-}\,\D^{i}\,\tldv_{i\,1}^{a} \Ra\,+\,\frac{1}{g}\,
\D_{j}\D^{j}\,\Le \Le\D^{i} U^{a b}\Ra\,\Le\D_{-}^{-1}\rho_{i}^{b}\Ra \Ra\,
+\,\Le \D_{+}\D_{-} + \D_{j}\D^{j}  \Ra\,\Le \D_{-}^{-2}\,j^{+}_{a\,1}\Ra\,=\,0\,,
\eeq
which we rewrite as
\beq\label{App91}
\Box\,\Le \D^{i}\,\tldv_{i\,1}^{a} \Ra\,=\,-\,\frac{1}{g}\,
\D_{j}\D^{j}\,\Le \Le\D^{i} U^{a b}\Ra\,\rho_{i}^{b}\Ra\,-\,
\Le\,\D_{+}\,j^{+}_{a\,1}\,+\,\D_{j}\D^{j}\D_{-}^{-1}\,j^{+}_{a\,1}\,\Ra\,.
\eeq
Therefore, with NLO precision, the answer is 
\beq\label{App10}
\tldv_{i\,1}^{a}\,=\,-\,\frac{1}{g}\,\Box^{-1}\,\D_{i}\,\Le\, \Le\D^{j} U^{a b}\Ra\,\rho_{j}^{b}\,\Ra\,-\,
\Box^{-1}\,\Le \,\D_{i}\D_{-}^{-1}\,j^{+}_{a\,1}\Ra\,.
\eeq

\item
 For the last equation, \eq{EA10}, 
\beq\label{App11}
2\Le D_{+}\,\Le \D_{-} \textsl{v}_{i} \Ra\,\Ra_{a}\,-\,\Le\,D_{i}\,\Le \D_{-} \textsl{v}_{+} \Ra\,\Ra_{a}\,+\,\Le D^{j}G_{ji}\Ra_{a}\,=\,0\,,
\eeq
at NLO we write
\beq\label{App101}
2\,\D_{+}\,\D_{-}\,\textsl{v}^{a}_{i\,1}\,+\,2\D_{+}\,\D_{-}\,\tldv^{a}_{i\,1}\,+\,\D_{j}^{2}\,\textsl{v}^{a}_{i\,1}\,+\,
\Le\,\D_{j}^{2}\,\tldv^{a}_{i\,1}\,-\,\D_{i}\,\D^{j}\,\tldv^{a}_{j\,1}\,\Ra\,+\,\D^{j}\,F_{j\,i}^{a}\,-\,\D_{i}\,\D_{-}\,\textsl{v}^{a}_{+\,1}\,=\,0\,,
\eeq
with $\D^{j}\,F_{j\,i}^{a}\,$ function as remaining NLO part of $\,D^{j}\,G_{j\,i}\,$ which depends on $\rho_{i}$ and $A_{+}$ fields only.
Therefore 
\beq\label{App12}
\textsl{v}^{a}_{i\,1}\,=\,-\,\Box^{-1}\,\Le\,\D^{j}\,F_{j\,i}^{a}\,\Ra\,
\eeq
with the same \eq{App91} for the $\tldv^{a}_{i\,1}$ function. The complete NLO correction to $\textsl{v}^{a}_{i}$ function reads as
\beq\label{App13}
\textsl{V}_{i\,1}^{a}=\textsl{v}_{i\,1}^{a}(x_{\bot}, x^{-}, x^{+})+
\tilde{\textsl{v}}_{i\,1}^{a}(x_{\bot}, x^{-}, x^{+})=-\Box^{-1}\,\Le\,\D^{j}\,F_{j\,i}^{a}+
\frac{1}{g}\,\D_{i}\Le\Le\D^{j} U^{a b}\Ra\rho_{j}^{b}\Ra+\D_{i}\D_{-}^{-1}\,j^{+}_{a\,1}\Ra\,.
\eeq
\end{enumerate}


\newpage
\begin{thebibliography}{99}

\bibitem{Gribov}
V. N. Gribov, Sov. Phys. JETP 26 (1968) 414.	

\bibitem{LipatovEff}
L.~N.~Lipatov,
  Nucl. Phys. B {\bf 452}, 369 (1995); Phys. Rept.  {\bf 286}, 131 (1997);
  Subnucl. Ser.  {\bf 49}, 131 (2013);
  Int. J. Mod. Phys. Conf. Ser.  {\bf 39}, 1560082 (2015);
	Int. J. Mod. Phys. A {\bf 31}, no. 28/29, 1645011 (2016);
	EPJ Web Conf.  {\bf 125}, 01010 (2016).
	
\bibitem{BFKL}	
L.~N.~Lipatov,
  Sov.\ J.\ Nucl.\ Phys.\  {\bf 23},  338 (1976)
  [Yad.\ Fiz.\  {\bf 23} (1976) 642];
%
  E.~A.~Kuraev, L.~N.~Lipatov and V.~S.~Fadin,
  Sov.\ Phys.\ JETP {\bf 45},  199 (1977)
  [Zh.\ Eksp.\ Teor.\ Fiz.\  {\bf 72},  377 (1977)];
%
  I.~I.~Balitsky and L.~N.~Lipatov,
  Sov.\ J.\ Nucl.\ Phys.\  {\bf 28},  822 (1978)
  [Yad.\ Fiz.\  {\bf 28},  1597 (1978)].	
	
\bibitem{Venug}
L.~McLerran and R.~Venugopalan, {\it{Phys.\ Rev.}}\ {\bf{D49}} (1994), 2233;
\ {\bf{D49}} (1994), 3352.

\bibitem{Kovner}
J.~Jalilian-Marian, A.~Kovner, L.~McLerran and H.~Weigert, 
{\it Phys.Rev.} {\bf D55},
5414 (1997);
J.~Jalilian-Marian, A.~Kovner, A.~Leonidov and H.~Weigert,
  Nucl.\ Phys.\ B {\bf 504}, 415 (1997);
	J.~Jalilian-Marian, A.~Kovner, A.~Leonidov and H.~Weigert,
  Phys.\ Rev.\ D {\bf 59}, 014014 (1998);
  J.~Jalilian-Marian, A.~Kovner and H.~Weigert,
  Phys.\ Rev.\ D {\bf 59}, 014015 (1998);
	E.~Iancu, A.~Leonidov and L.~D.~McLerran,
  Nucl.\ Phys.\ A {\bf 692}, 583 (2001);
	E.~Iancu, A.~Leonidov and L.~D.~McLerran,
  Phys.\ Lett.\ B {\bf 510}, 133 (2001);
	E.~Ferreiro, E.~Iancu, A.~Leonidov and L.~McLerran,
  Nucl.\ Phys.\ A {\bf 703}, 489 (2002).

\bibitem{GLR}
L. V. Gribov, E. M. Levin and M. G. Ryskin, Phys.Rep. 100, 1 (1983).

\bibitem{Hatta1}
Y.~Hatta,
  Nucl.\ Phys.\ A {\bf 768}, 222 (2006);
 Y.~Hatta,
  Nucl.\ Phys.\ A {\bf 781}, 104 (2007).

\bibitem{MaMi}	
Y. M. Makeenko and A. A. Migdal,  Physics Letters 88B, 135-137 (1979).	
	
	
\bibitem{Heinz}
U.~Heinz, Phys.\ Rev.\ Lett.\ {\bf 51} (1983) 351; Phys.\ Lett.\ {\bf B144} (1984) 228;
Annals Phys.\ {\bf 161} (1985) 48; Annals Phys.\ {\bf 168} (1986) 148;
Z.\ Phys.\ {\bf C38} (1988) 203; Physica {\bf A158} (1989) 111.

\bibitem{VenPlas} 
J.~Jalilian-Marian, S.~Jeon, R.~Venugopalan and J.~Wirstam,
  Phys.\ Rev.\ D {\bf 62}, 045020 (2000).

\bibitem{VenPlas1} 
  J.~Jalilian-Marian, S.~Jeon and R.~Venugopalan,
  Phys.\ Rev.\ D {\bf 63}, 036004 (2001).
	
	\bibitem{Iancu} 
  J.~P.~Blaizot and E.~Iancu,
  Phys.\ Rept.\  {\bf 359}, 355 (2002)

\bibitem{Plas1}
D.~F.~Litim and C.~Manuel,
  Phys.\ Rept.\  {\bf 364}, 451 (2002).
	
	


	
\bibitem{BK}
I. Balitsky, Nucl. Phys. B463 (1996) 99; Y. V. Kovchegov, Phys. Rev. D60 (1999) 034008; Phys. Rev. D61 (2000) 074018.	

\bibitem{Braun}  
M.~A.~Braun,  Phys.\ Lett.\ B {\bf 483} (2000) 115; Eur.\ Phys.\ J.\ C {\bf 33} (2004) 113.

\bibitem{Braun1}
S.~Bondarenko and M.~A.~Braun,
  Nucl.\ Phys.\ A {\bf 799}, 151 (2008).
	
\bibitem{Higgs} 
J.~Bartels, S.~Bondarenko, K.~Kutak and L.~Motyka,
  Phys.\ Rev.\ D {\bf 73}, 093004 (2006)


\bibitem{Bom}  
S.~Bondarenko and L.~Motyka,  Phys.\ Rev.\  D {\bf 75}, (2007) 114015.

\bibitem{Bond}
S.~Bondarenko,
  Phys.\ Lett.\ B {\bf 665}, 72 (2008); Nucl.\ Phys.\ A {\bf 853}, 71 (2011).
	
\bibitem{OldRFT}  
D.~Amati, L.~Caneschi and R.~Jengo,  Nucl.\ Phys.\ B {\bf 101} (1975) 397;
 R.~Jengo, Nucl.\ Phys.\ B {\bf 108} (1976) 447; M.~Ciafaloni,  Nucl.\ Phys.\ B {\bf 146} (1978) 427.	

 \bibitem{NewRFT1}
E.~Levin and A.~Prygarin,
  Eur.\ Phys.\ J.\ C {\bf 53}, 385 (2008);
S.~Bondarenko,
  Eur.\ Phys.\ J.\ C {\bf 71}, 1587 (2011); 
	S.~Bondarenko, L.~Horwitz, J.~Levitan and A.~Yahalom,
  Nucl.\ Phys.\ A {\bf 912}, 49 (2013).
	

	


\bibitem{EffAct}
L.~N.~Lipatov,
  Nucl.\ Phys.\ Proc.\ Suppl.\  {\bf 99A}, 175 (2001);
	 M.~A.~Braun and M.~I.~Vyazovsky,
  Eur.\ Phys.\ J.\ C {\bf 51}, 103 (2007);
	 M.~A.~Braun, M.~Y.~Salykin and M.~I.~Vyazovsky,
  Eur.\ Phys.\ J.\ C {\bf 65}, 385 (2010);
	M.~A.~Braun, L.~N.~Lipatov, M.~Y.~Salykin and M.~I.~Vyazovsky,
  Eur.\ Phys.\ J.\ C {\bf 71}, 1639 (2011);
	 M.~A.~Braun, M.~Y.~Salykin and M.~I.~Vyazovsky,
  Eur.\ Phys.\ J.\ C {\bf 72}, 1864 (2012);
	M.~Hentschinski and A.~Sabio Vera,
  Phys.\ Rev.\ D {\bf 85}, 056006 (2012);
	M.~A.~Braun, M.~Y.~Salykin, S.~S.~Pozdnyakov and M.~I.~Vyazovsky,
  Eur.\ Phys.\ J.\ C {\bf 72}, 2223 (2012);
	J.~Bartels, L.~N.~Lipatov and G.~P.~Vacca,
  Phys.\ Rev.\ D {\bf 86}, 105045 (2012);
	M.~A.~Braun, S.~S.~Pozdnyakov, M.~Y.~Salykin and M.~I.~Vyazovsky,
  Eur.\ Phys.\ J.\ C {\bf 73}, no. 9, 2572 (2013);
	G.~Chachamis, M.~Hentschinski, J.~D.~Madrigal Martínez and A.~Sabio Vera,
  Phys.\ Part.\ Nucl.\  {\bf 45}, no. 4, 788 (2014).
	
\bibitem{BKP}
J. Bartels, Nucl. Phys. B 175 (1980) 365;
J. Kwiecinski, M. Praszalowicz, Phys. Lett. B 94 (1980) 413.

\bibitem{BKP1}	
J.~Bartels, V.~S.~Fadin, L.~N.~Lipatov and G.~P.~Vacca,
  Nucl.\ Phys.\ B {\bf 867}, 827 (2013).
	
\bibitem{Hatta2}
I.~Balitsky,
  Phys.\ Rev.\ D {\bf 72}, 074027 (2005);
	Y.~Hatta, E.~Iancu, L.~McLerran, A.~Stasto and D.~N.~Triantafyllopoulos,
  Nucl.\ Phys.\ A {\bf 764}, 423 (2006).	
	
\bibitem{Bal}
I.~Balitsky,
  Phys.\ Rev.\ D {\bf 60}, 014020 (1999);
In *Shifman, M. (ed.): At the frontier of particle physics, vol. 2* 1237-1342;
Nucl.\ Phys.\ B {\bf 629}, 290 (2002).		
			
	
\end{thebibliography}
\end{document}